\documentclass[]{aastex631}

\graphicspath{{./}{figures/}}
\usepackage{hyperref}
\usepackage{float}
\usepackage{graphicx}
\usepackage{xcolor}
\usepackage{colortbl}
\usepackage{color}
\begin{document}

\title{Measuring the Intensity  of the Interstellar Radiation Field with Ultra-high Energy $\gamma$ ray Spectra}

\author[0000-0002-1188-7503]{Nayantara Gupta}
\affiliation{Raman Research Institute \\
C. V. Raman Avenue, 5th Cross Road, Sadashivanagar, Bengaluru, Karnataka 560080, India}
\email{nayan@rri.res.in}

\begin{abstract}
Our understanding of the intensity distribution of the interstellar radiation background is based on the observational data from IRAS, COBE-FIRAS and Planck.  The intensity of this radiation field increases rapidly towards the Galactic plane and is the highest near the Galactic centre due to the high density of stars and dust. However,  a precise determination of the variations of this radiation field with spatial and angular coordinates is not feasible observationally. We explore how future studies of gamma-ray spectra from numerous ultra-high-energy 
gamma-ray sources can indirectly probe variations in the interstellar radiation field's intensity across different distances from the Galactic centre and across Galactic latitudes and longitudes.  This study is crucial for making self-consistent predictions of high energy gamma-ray fluxes from Galactic sources detected by observatories like LHAASO, Tibet AS$_{\gamma}$ and the next-generation gamma-ray telescopes.
 
\end{abstract}

\keywords{High-energy astrophysics(739) -- Gamma rays (637) -- Gamma ray sources (633)}

\section{Introduction} \label{sec:intro}
The emission by stars and the scattering, absorption and reemission of the absorbed starlight by interstellar dust result in the formation of ISRF. The relative distribution of stellar emissivity and dust opacity varies throughout the Galaxy, complicating the estimation of optical depth.
The intensity of ISRF varies with the radial distance (r) from the Galactic centre and distance (z) from the Galactic plane, and it also has angular dependence on longitude (l) and latitude (b).  Since it is not directly measurable, there have been many attempts to model it in different ways. Some of the early models are discussed in these papers, \citet{1974csgr.conf..229C}; \citet{1976ApJ...207L..49F}; \citet{1976A&A....52...69P}; \citet{1976ApJ...208..893S}; \citet{1977A&A....59..233B}. The model by \citet{1983A&A...128..212M} has been widely used to get the ISRF covering far-infrared to UV frequencies at some locations in the Galaxy. Further investigations were carried out to model the ISRF by \citet{1985A&A...145..391B}; and \citet{1991JPhG...17..987C}. Although there were many attempts, there was no self-consistent model to describe the ISRF. The three-dimensional model by \citet{2012A&A...545A..39R} uses the SKY model discussed in  \citet{1992ApJS...83..111W}, \citet{1993AJ....105.1860C} and \citet{1994AJ....107..582C} for a self-consistent estimation of ISRF intensities in the solar neighbourhood. This topic is inherently complex due to the lack of a definitive model to describe the structure of the Galaxy, for a detailed discussion on this, see \citet{2000ApJ...537..763S}, \citet{Porter_2008}, \citet{Porter_2017}, \citet{Popescu_2017} and \citet{Vernetto_2016}. The observational data from IRAS \citep{Miville_Deschenes_2005}, COBE-FIRAS (\citet{1991ApJ...381..200W}, \citet{Reach_1995}, \citet{1999ApJ...524..867F}), COBE-DIRBE \citep{Arendt_1998} and Planck (\citet{2014}, \citet{Fanciullo_2015} ) have been used to test the model predictions of ISRF distribution.
\par
Ground-based detectors like \href{https://www.mpi-hd.mpg.de/hfm/HESS/}{H.E.S.S} (High Energy Stereoscopic System), \href{https://magic.mpp.mpg.de/}{MAGIC} (Major Atmospheric Gamma Imaging Cherenkov) telescope, \href{https://www.icrr.u-tokyo.ac.jp/em/}{Tibet}, \href{https://www.hawc-observatory.org/}{HAWC} (High-Altitude Water Cherenkov) gamma-ray observatories,  \href{http://english.ihep.cas.cn/lhaaso/}{ LHAASO} (Large High Altitude Air Shower Observatory) and space-based detectors  \href{http://www-glast.stanford.edu/}{Fermi LAT}, \href{http://agile.rm.iasf.cnr.it/}{AGILE} have observed many ultra-high-energy (UHE) gamma-ray sources. With the detection of gamma rays with energies of several hundreds of TeV to PeV in our Galaxy, UHE gamma-ray astronomy has opened a new window to explore the most energetic phenomena and their underlying physical mechanisms \citep{annurev:/content/journals/10.1146/annurev-nucl-112822-025357}. The UHE gamma rays interact with the interstellar radiation field (ISRF) and the cosmic microwave background (CMB) to produce pairs of electrons and positrons. As a result of these interactions, the spectrum of the UHE gamma rays is attenuated, and a change in slope appears in the injected spectrum. The threshold energy condition of the pair production interactions determines the energy where the slope of the original spectrum deviates. Hence, it is a uniquely distinguishable feature to identify the effect of the optical depth due to the background radiations. The optical depth depends on the intensity of ISRF and CMB radiations and the distance of the source from the observer. The intensity of ISRF changes along the line of sight of the observer; hence, it is important to have precise knowledge of the spatial and angular distribution of ISRF. 
\par
The UHE gamma rays may have leptonic origin; in this case, the very high energy electrons from pulsars or supernova remnants lose energy by inverse Compton mechanism and secondary gamma rays are produced. They may also be made in hadronic interactions; in this case, the very high energy cosmic ray protons from supernova remnants may interact with molecular clouds to create the gamma rays. A power law or a log parabola spectrum of UHE gamma rays may be produced depending on the spectral shape of the parent cosmic ray spectrum. Due to the absorption of  UHE gamma rays by ISRF and CMB,  deviations in the spectral shape from simple power-law spectrum are expected near a few hundred TeV and 2 PeV, respectively. Thus, irrespective of the underlying mechanisms of UHE gamma-ray production, the signature of attenuation by the background radiations would be visible in the spectrum. We emphasise the fact that even if we do not know whether the origin of the UHE gamma-rays is leptonic or hadronic, the knowledge of the extent of deviation from a power law spectrum near a few hundred TeV could be used to estimate the effect of the optical depth, which could be useful to measure the ISRF along the line of sight of the observer.

\par
We discuss how detecting the spectra from a large number of UHE gamma-ray sources located at different r, z, l, and b and modelling their spectral deviations from simple power law spectra (intrinsic/source spectra) could indirectly explore the spatial and angular dependence of ISRF.
 In this work, we assume that the ISRF does not vary with l and b for simplification; however, in principle, this study could be extended to account for the angular variation in ISRF. 
 In section 2, we show the variation in optical depth with distance from the observer when the source is near the Galactic plane, as the attenuation in the UHE gamma-ray spectra due to pair production with ISRF is most significant in this region. In section 3, we give a list of some important UHE gamma-ray sources which have been included in this study. Our results are mentioned in Section 4 and discussed in Section 5.

\section{Optical Depth due to ISRF and CMB}
The radiation field in our Galaxy has contributions from stars, dust, CMB and extragalactic background light (EBL). In our calculation, we have not included the contribution from EBL as it is not as important as the other components.   
\citet{2012A&A...545A..39R} presented a self-consistent three-dimensional radiative transfer model of the stellar and dust emission in the Milky Way. The factorised form of dust distribution given in Eqn.(1) of their paper is
\begin{equation}
\rho(r,z,s)=\rho_d(s) \exp\Big[-\frac{r}{h_d}-\frac{|z|}{z_d}\Big]
\end{equation}
where $r$, $z$ are cylindrical coordinates and $s$ denotes the spectral class; $h_d=$3.5 kpc, $z_d$=100 pc and $\rho_d=10^{-25}$ gm/cm$^{-3}$. The intensity distribution of ISRF with longitude and latitude is shown in Fig.2. to Fig.5. of their paper and compared with the observational data from the GLIMPSE surveys carried out using the  InfraRed Array Camera (IRAC), MIPSGAL surveys carried out using the Multiband Imaging Photometer for Spitzer (MIPS) and Improved Reprocessing of the IRAS Survey (IRIS). We will discuss the features of this intensity distribution later. We focus on the factorised form of the radial distribution.
\par
The dust emission model by \citet{Misiriotis_2006} has been used by \citet{Vernetto_2016}, which has a few parameters for interstellar dust's spatial and temperature distributions. This dust emission model is consistent with the main features of the spectral data of COBE-FIRAS (\citet{1991ApJ...381..200W}, \citet{Reach_1995}, \citet{1999ApJ...524..867F}) and the sky maps of COBE-DIRBE \citep{Arendt_1998}. The dust emission has cylindrical symmetry, mostly contained in the region r$<$10kpc and z$<$2kpc. 
\begin{equation}
\rho_{c,w}(r,z)=\rho^{o}_{c,w} exp\Big(-\frac{r}{R_{c,w}}-\frac{|z|}{Z_{c,w}}\Big).
\label{dust_dist}
\end{equation}

In Eqn(\ref{dust_dist})  $\rho_c(\rho_w)$ is the mass density of the cold (warm) dust, r and z are cylindrical coordinates in the Galaxy.
The parameters $R_c, Z_c$ and $R_w, Z_w$ represent the scale length and height of the cold and warm dust distributions, respectively. The dust mass density for cold (warm) dust at the Galactic centre is denoted by $\rho^o_{c}(\rho^o_{w})$. The values of these parameters are given in
\citet{Misiriotis_2006} and Table 1 of \citet{Vernetto_2016}, $R_c= 5$ kpc, $Z_c=0.1$ kpc; $R_w=3.3$ kpc, $Z_w=0.09$ kpc. The contribution from warm dust is only 0.31$\%$. We have adopted this formulation in this study.  The factorised form allows us to constrain the radial dependence of ISRF
 with sources close to the Galactic plane, where $|z|\simeq0$.
\par
The distribution of the starlight energy density has been used from the model by \citet{1983A&A...128..212M}. It is represented by a cylindrically symmetric and factorised form in r and z in Eq.(22) of \citet{Vernetto_2016}. 
\begin{equation}
u_{\lambda}(r,z)=u^{GC}_{\lambda} e^{-r/R_s-|z|/Z_s}.
\label{star_l}
\end{equation}

In Eqn.(\ref{star_l}) the parameter $R_s$=2.17 kpc, (4.07 kpc) for r$<$8 kpc, (r$>$8 kpc) and $Z_s$=7.22 kpc.
Fig.11. of their paper shows the sum of the stellar and dust energy spectra at the Galactic centre and various radial distances from it. We have used 
the energy distribution of ISRF photon number density at the Galactic centre from Fig.11. of \citet{Vernetto_2016}, and subsequently used the factorised form given in Eqn. (\ref{dust_dist}) and Eqn. (\ref{star_l}) to get the distribution of ISRF with energy and radial distance.

\par
\begin{figure}[!h]
 \centering
 \includegraphics[scale=0.4]{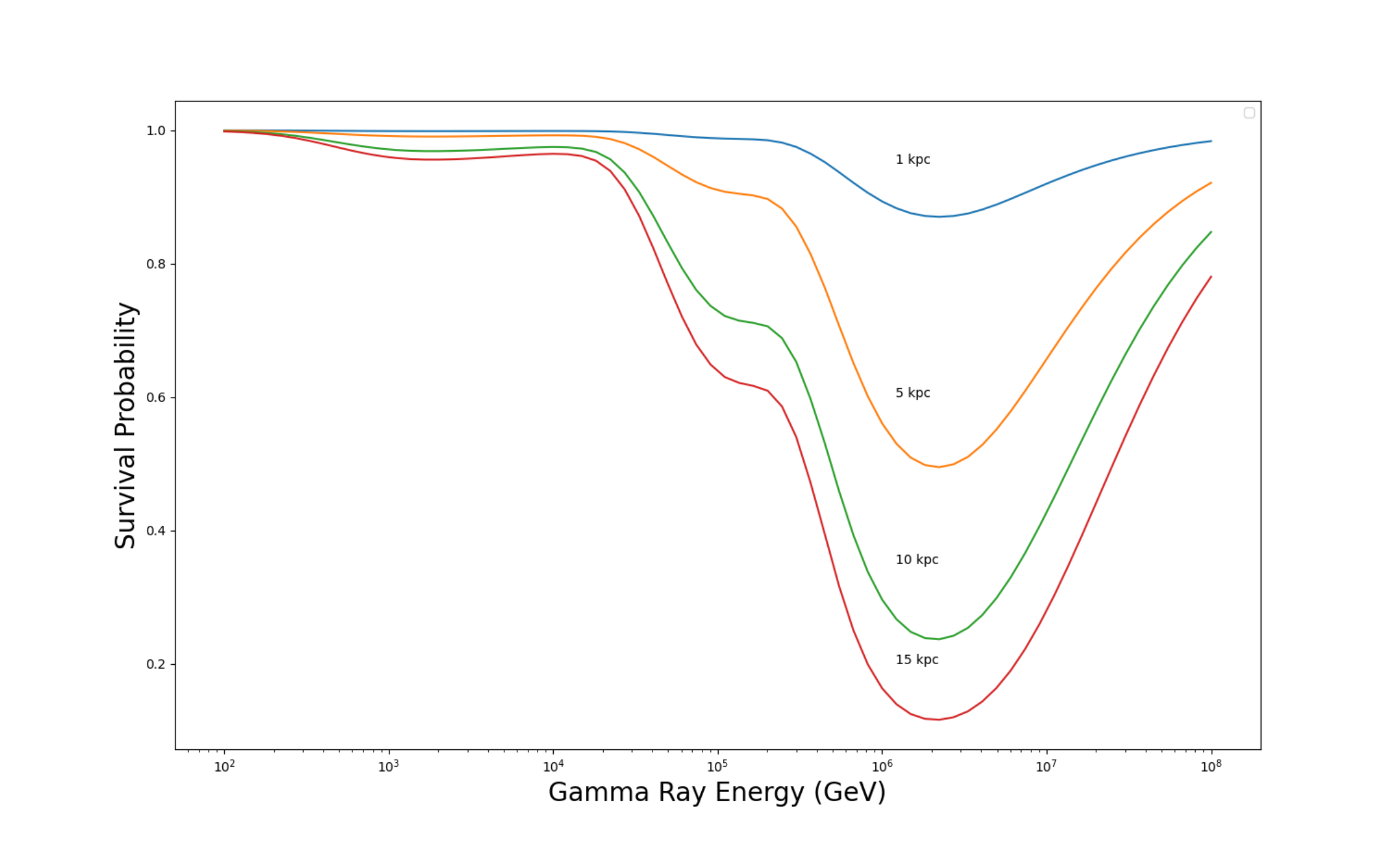}
 \includegraphics[scale=0.4]{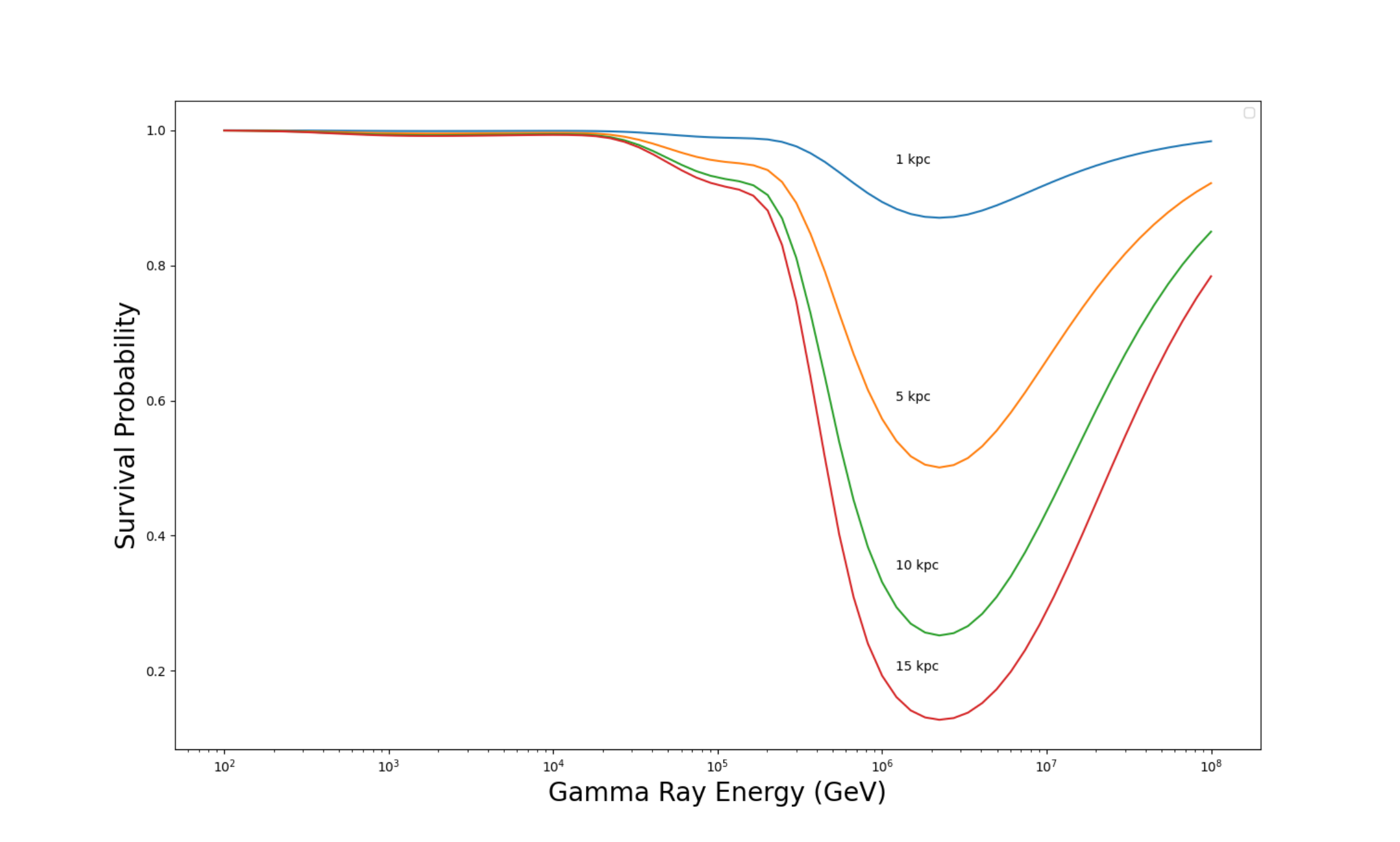}
 \caption{Survival probability of high energy gamma rays plotted for distances of the source one kpc, 5 kpc, 10 kpc and 15 kpc shown by curves from top to bottom; longitude 0$^{\circ}$ (top figure) and 100$^{\circ}$ (bottom figure); latitude 0$^{\circ}$ calculated using the dust emission and starlight energy density distribution given in Eqn. (\ref{dust_dist}) and Eqn. (\ref{star_l}), with $R_c$=5 kpc.}
 \label{fig: opt_ver_rad}
\end{figure}

After incorporating this ISRF, the optical depth of high-energy gamma rays due to pair production in background radiation has been calculated numerically following the formalism discussed by \citet{PhysRev.155.1404}. The inverse of the mean free path length of $\gamma \gamma$ interactions between a high energy gamma ray of energy $E_t$ and low energy background radiation of energy $\epsilon$, with threshold energy of interactions $\epsilon_{th}$ is 
\begin{equation}
l^{-1}_{\gamma \gamma}(E_t)=\frac{1}{2} \int d(cos{\theta}) (1-cos{\theta}) \int^{\infty}_{\epsilon_{th} }  d\epsilon \frac{dn(\epsilon,r,z)}{d\epsilon} \sigma_{\gamma\gamma}. 
\end{equation}
The number density of photons in the radiation background $n(\epsilon,r,z)$ is a function of energy ($\epsilon$), radial distance ($r$) and distance from the Galactic plane ($z$).
The effect of angular variation in ISRF is important, and it can be factorised and studied in detail in the future.
The cross-section of interactions $\sigma_{\gamma \gamma}$ is a function of $E_t$, $\epsilon$ and the angle between two-photon directions $\theta$.
The threshold energy of interactions
\begin{equation}
\epsilon_{th}=\frac{2 (m_e c^2)^2}{E_t(1-cos{\theta})}.
\end{equation}
The optical depth for the gamma rays of energy $E_t$ travelling through a distance $d$ from the source to the observer is given by
\begin{equation}
\tau(E_t)=d   \  l^{-1}_{\gamma \gamma} (E_t)
\end{equation}
The survival probability of high-energy gamma rays of energy $E_t$ is $\exp(-\tau(E_t))$. It has been plotted in Fig \ref{fig: opt_ver_rad}, after varying the distance of the UHE gamma-ray source from the observer for longitude 0$^{\circ}$ and 100$^{\circ}$ respectively and latitude 0$^{\circ}$.  Subsequently, we have calculated the attenuated spectra of UHE gamma-ray sources located near the  Galactic plane.  We have included CMB while calculating the optical depth of the UHE gamma rays.
Gamma rays' intrinsic or source spectrum is assumed to be a power law spectrum $ I_ 0(E_t)$. The attenuated spectrum is $I(E_t)$.
\begin{equation}
I(E_t)= I_0(E_t) \exp(-\tau(E_t))
\label{spec}
\end{equation}
The gamma rays, which have energy near 100 TeV, interact mainly with the dust emission and CMB.
We have constrained the scale length ($R_c$) in the radial component of the ISRF $\exp(-r/R_c)$ in Eqn. (\ref{dust_dist}), with the uncertainty in the observed UHE gamma-ray flux from different UHE gamma-ray sources listed in Table. \ref{tab:params}.  Note that as the value of the scale length vanishes, $R_c \rightarrow 0$, the radiation field due to dust emission vanishes for the sources near the Galactic plane and as a result the optical depth due to dust emission also vanishes, hence in this case the observed UHE spectrum will have the attenuation due to CMB.
 
\section{Gamma Ray Spectra of Ultra-high Energy Gamma Ray Sources:}
The first catalogue \citet{Cao_2024} was compiled using 508 days of data collected by the Water Cherenkov Detector Array (WCDA) from March 2021 to September 2022 and 933 days of data recorded by the  Kilometer Squared Array (KM2A) from January 2020 to September 2022. This catalogue contains 90 sources with an extended size smaller than 2$^{\circ}$ and a significance of detection more than 5$\sigma$. Below, we have discussed the extended emission from several LHAASO sources to explain how their spectral shapes can be used to probe the intensity distribution of the ISRF.

\subsection{1LHAASO J1843-0335u}
WCDA and KM2A of LHAASO have detected this source, which is associated with HESS J1843-033. SNR G28.6-0.1 may be its counterpart (\citet{Cao_2024}, \citet{zhang2024attenuationlhaasopevatronsinterstellar}), which is located at a distance of 9.6 kpc and its longitude and latitude are 28.84$^{\circ}$ and 0.09$^{\circ}$, respectively. The gamma-ray spectrum detected by KM2A has been fitted with a power law by \citet{Cao_2024}. The values of the parameters of this fit and the errors in these values have been taken from their paper. We have plotted this in Fig \ref{fig: J1843}. The green dashed line shows the power law fit. The blue solid line shows the attenuated spectrum.
\subsection{1LHAASO J1914+1150u}
This source is associated with HAWC detected source 2HWC J1914+117* and may be associated with PSR J1915+1150 ( \citet{Cao_2024}, \citet{zhang2024attenuationlhaasopevatronsinterstellar}), which is located at a distance of 14.01 kpc having longitude and latitude are 46.13$^{\circ}$ and 0.26$^{\circ}$ respectively. Its gamma-ray spectrum has been plotted in Fig \ref{fig: J1914}. 
\subsection{1LHAASO J1929+1846u*}
This source is possibly associated with SNR G054.1+00.3. It is located at a distance of 7 kpc, and its longitude and latitude are 53.88$^{\circ}$ and 0.45$^{\circ}$ respectively  (\citet{Cao_2024}, \citet{zhang2024attenuationlhaasopevatronsinterstellar}).
\subsection{1LHAASO J1959+1129u}
Lastly, we have chosen this source from \citet{Cao_2024}; it has also been studied by \citet{zhang2024attenuationlhaasopevatronsinterstellar}. Its possible counterpart is a low-mass X-ray binary located at a distance of 9.4 kpc, and its longitude and latitude are 51.1$^{\circ}$ and -9.42$^{\circ}$ respectively.
\subsection{Deviation from a Simple Power law spectrum}
We have taken some examples of UHE gamma-ray sources to study how the deviation in the spectral shape of the attenuated spectrum can be used to constrain the opacity of the medium along the line of sight and subsequently to estimate the intensity of ISRF.  Here, the source/intrinsic spectrum is assumed to be a power law spectrum. The black-shaded regions in Fig \ref {fig: J1843} to Fig \ref{fig: vary_rad_comp2} show the errors in the measurements by the LHAASO detector. In future, this region will become narrower as the precision in measurement increases, and it will also be possible to extend the measurement of the spectra to higher energies. This will help us to constrain the opacity along the line of sight with the observed UHE gamma-ray spectra from sources located at various distances and angular positions. The attenuated spectra should be contained within the black-shaded regions. We discuss this in more detail in the next section.

\section{Results}

\begin{figure}
 \includegraphics[scale=0.4]{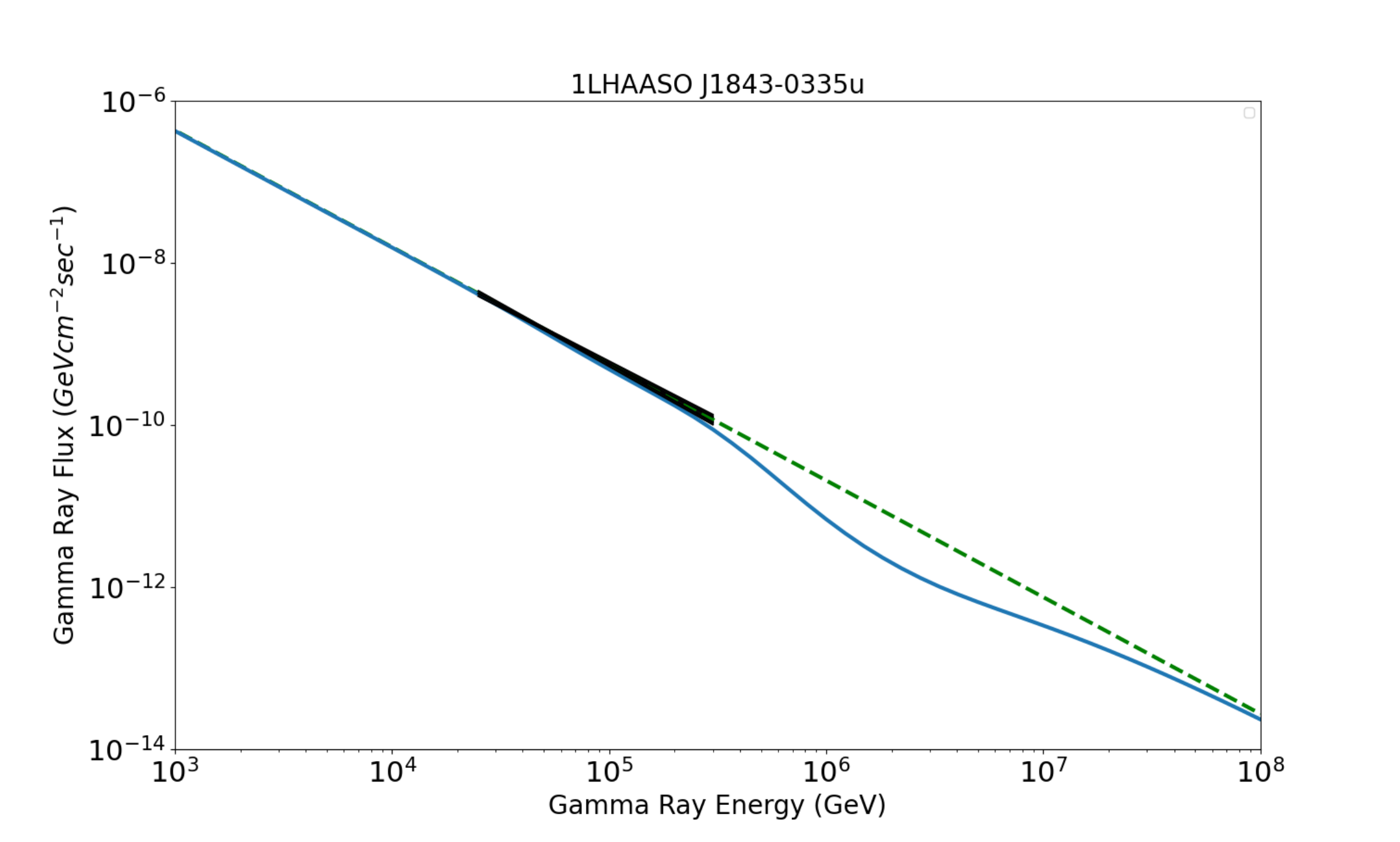}
 \caption{1LHAASO J1843-0335u: possible counterpart is SNR G28.6-0.1  located at a distance of 9.6 kpc, its longitude and latitude are  28.84$^{\circ}$ and 0.09$^{\circ}$ respectively. 
The black-shaded region shows the gamma-ray flux measured by LHAASO after including the measurement errors; the green dashed line shows the power law spectrum and the blue solid line shows the attenuated spectrum calculated with $R_c=4$ kpc.}
\label{fig: J1843}
\end{figure}

\begin{figure}[!h]
\centering
 \includegraphics[scale=0.4]{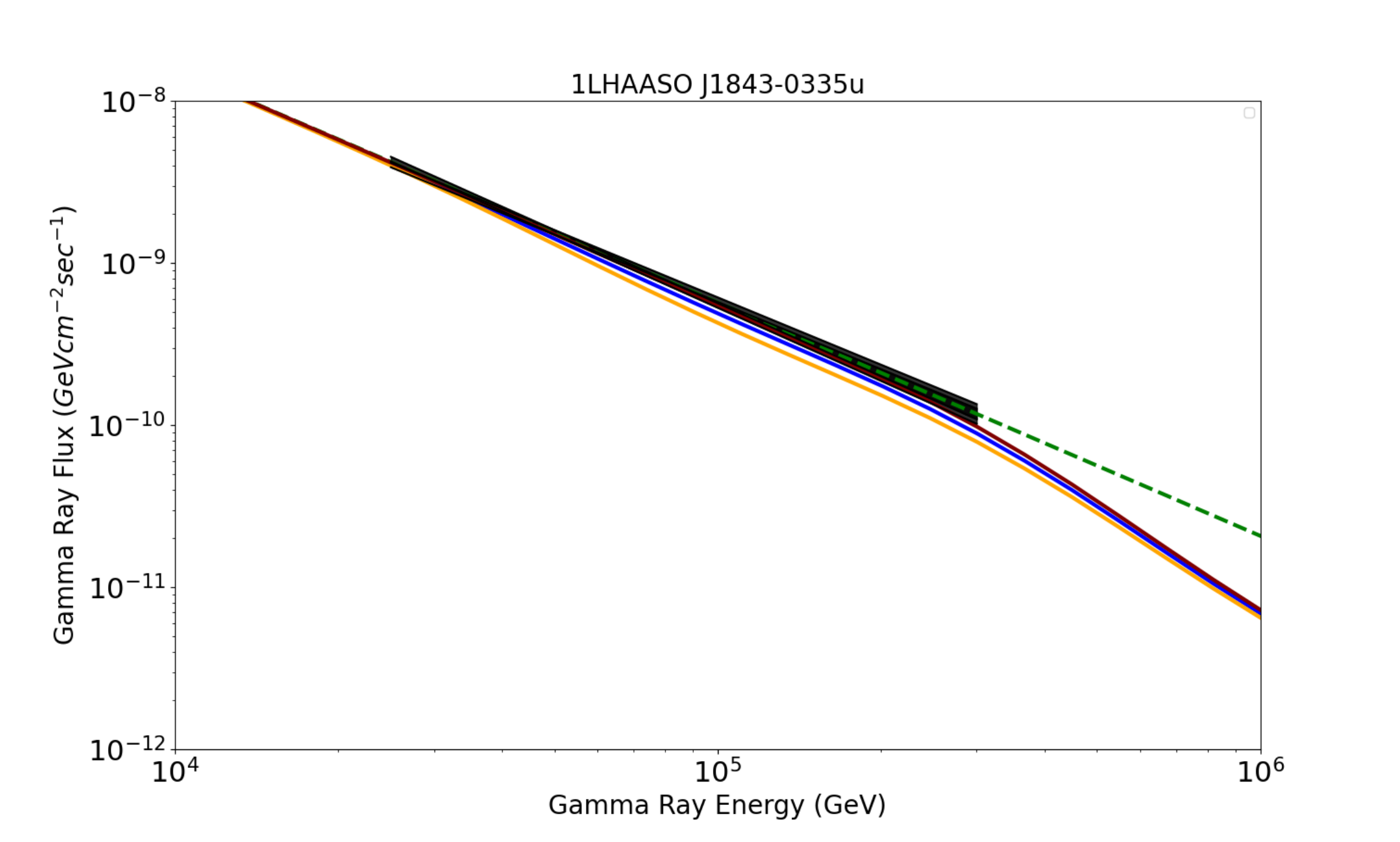}
 \caption{This is a zoomed-in version of Figure 2. We have assumed three different radial dependencies of ISRF ( maroon: $e^{-r/2}$, blue: $e^{-r/4}$ and orange: $e^{-r/8}$, which corresponds to $R_c=2$ kpc, 4 kpc and 8 kpc respectively.}
 \label{fig:vary_rad_comp1}
\end{figure}

In Fig \ref{fig: J1843}, we have plotted the UHE gamma-ray spectrum from 1LHAASO J1843-0335u located at longitude l=28.84$^{\circ}$ and latitude b=0.09$^{\circ}$. The black-shaded region shows the uncertainty in the measurement of the gamma-ray flux, the green dashed line shows the power law fit given by \citet{Cao_2024}, and the attenuated spectrum calculated in our work is shown with a blue line assuming the intrinsic or the source spectrum is represented by the green dashed line. We zoom into a smaller energy range to show the fit to the LHAASO data with the attenuated spectrum in Fig \ref {fig:vary_rad_comp1}. The radial dependency of ISRF has been varied to show the change in the attenuated spectrum.  We have considered the following values of $R_c$, 2 kpc, 4 kpc and 8 kpc. The maroon line corresponding to the radial dependency $e^{-r/2}$ of the ISRF is near the lower edge of the black-shaded region. In future, with more precise measurements of the UHE gamma-ray spectrum, the width of the black-shaded region will become narrower, and it will be possible to constrain the radial variation of ISRF more precisely for this longitudinal direction  28.84$^{\circ}$.

\begin{figure}[!h]
\centering
 \includegraphics[scale=0.4]{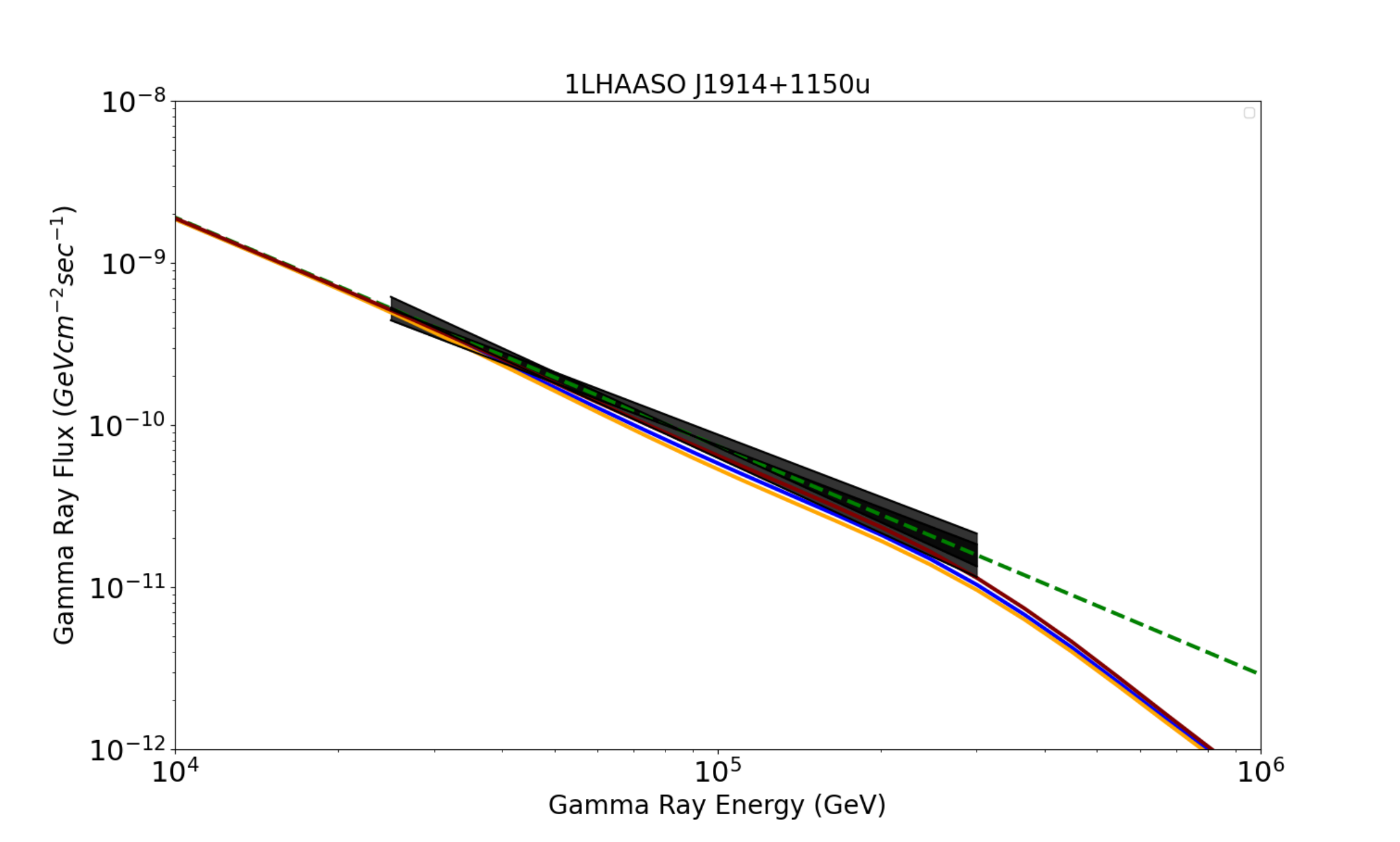}
 \caption{1LHAASO J1914+1150u: possible counterpart is PSR J1915+1150  located at a distance of 14.01 kpc, its longitude and latitude are  46.13$^{\circ}$ and 0.26$^{\circ}$ respectively. The black shaded region and green dashed line are similar to Fig \ref{fig: J1843}; the attenuated spectrum is calculated assuming different radial dependencies of ISRF (maroon: $e^{-r/4}$, blue: $e^{-r/6}$ and orange: $e^{-r/8}$, which corresponds to $R_c=4$ kpc, 6 kpc and 8 kpc respectively).}
 \label{fig: J1914}
\end{figure}

We have shown the UHE gamma-ray spectrum of another source, 1LHAASO  J1914+1150u, from the first LHAASO catalogue in Fig \ref{fig: J1914}. This source is also close to the Galactic plane but further away from us. In this case, the attenuated spectrum is calculated for three different values of $R_c$, 4 kpc, 6 kpc and 8 kpc. The attenuated spectrum corresponding to $ R_c=$4 kpc is near the edge of the black-shaded region.
\begin{figure}[!h]
\centering
 \includegraphics[scale=0.4]{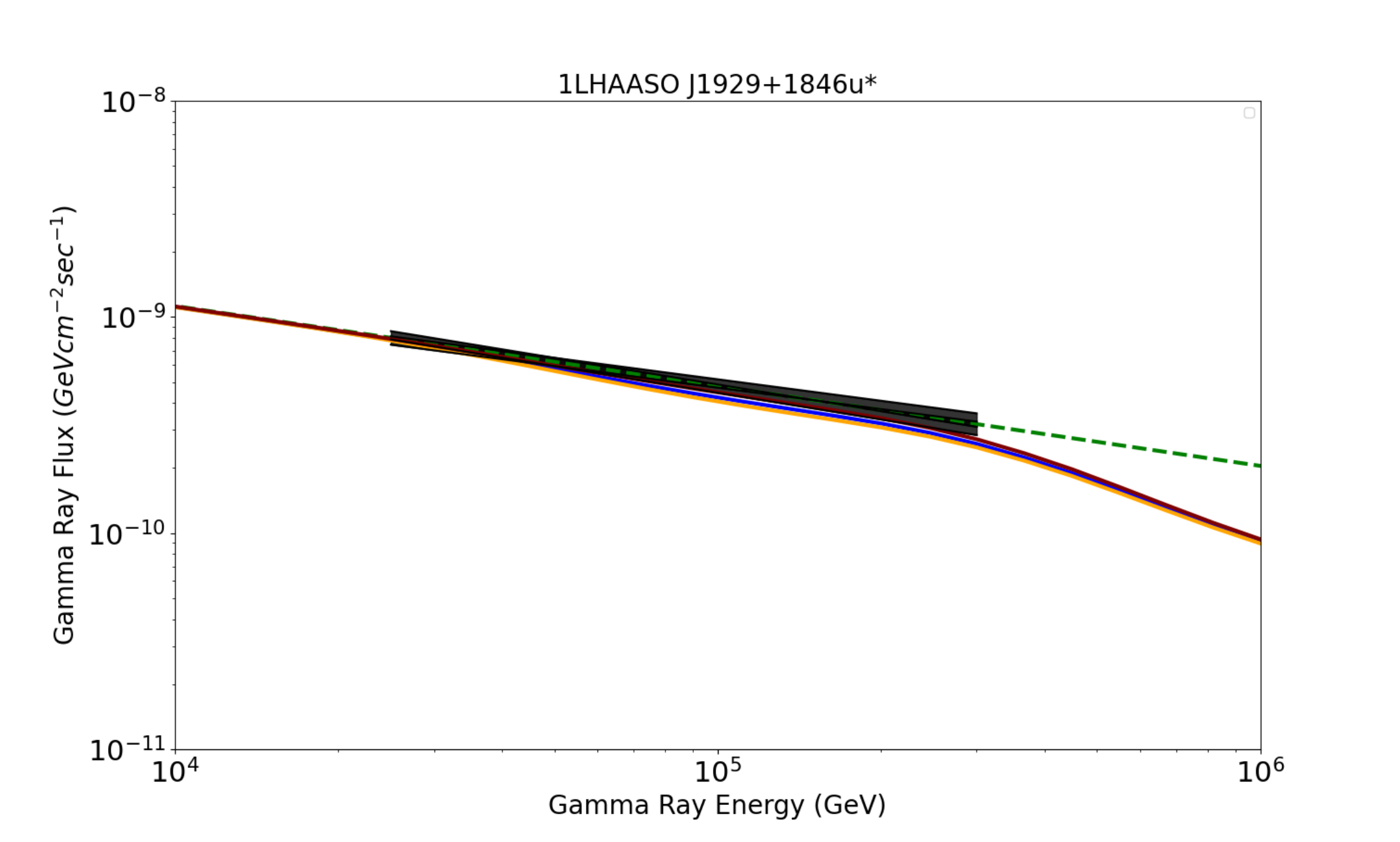}
 \caption{1LHAASO J1929+1846u*: possible counterpart is SNR G054.1+00.3  located at a distance of 7 kpc, its longitude and latitude are  53.88$^{\circ}$ and 0.45$^{\circ}$ respectively. The black shaded region and green dashed line are similar to Fig \ref{fig: J1843}; the attenuated spectrum is calculated assuming maroon: $R_c=4$ kpc, blue: $R_c=6$ kpc and orange: $R_c=8$ kpc similar to Fig \ref{fig: J1914}.}
 \label{fig: J1929}
\end{figure}
\begin{figure}[H]
\includegraphics[scale=0.4]{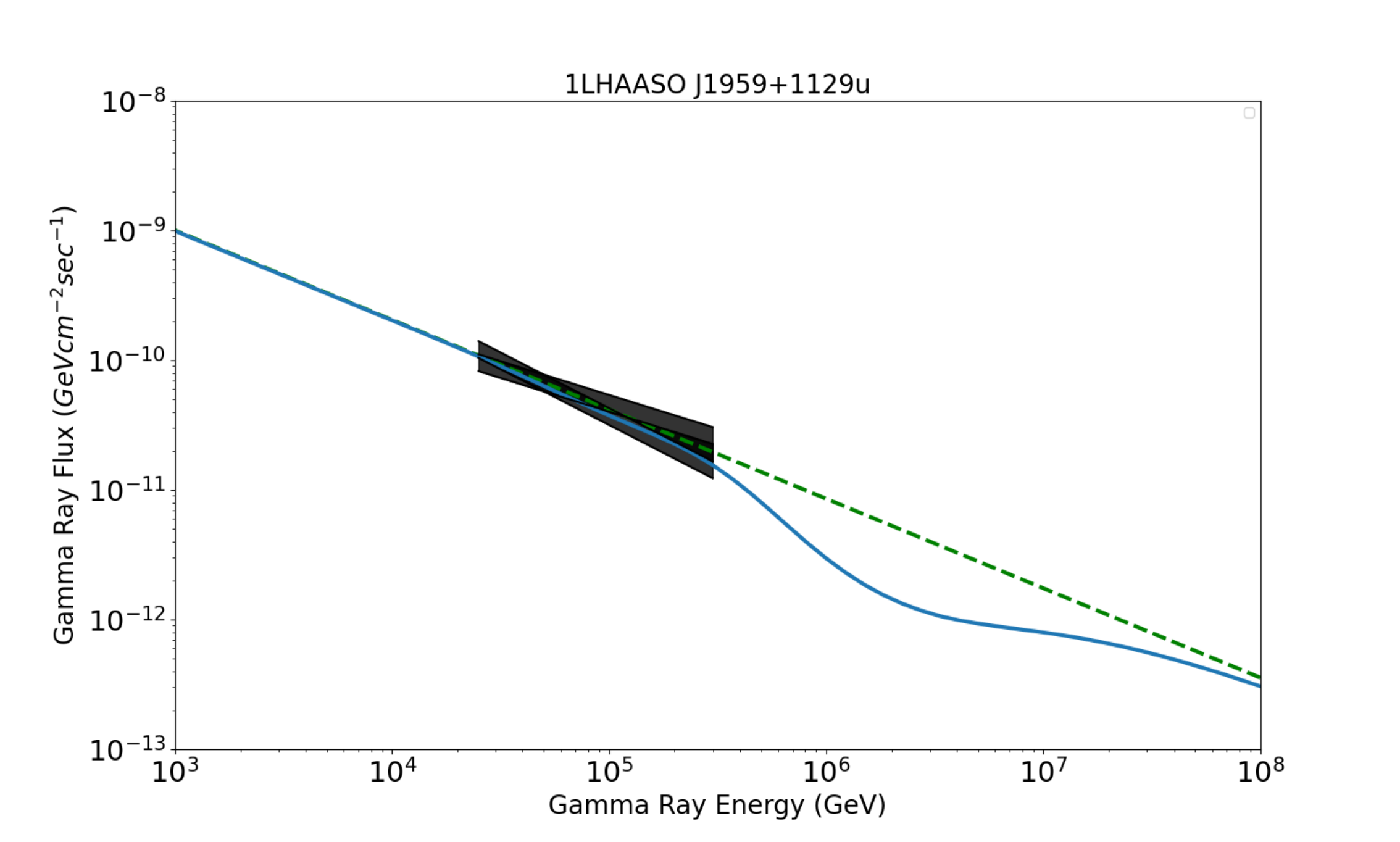}
\caption{1LHAASO J1959+1129u: possible counterpart is a low mass X-ray binary counterpart  4U 1957+11 located at a distance of 9.4 kpc, its longitude and latitude are  51.10$^{\circ}$ and -9.42$^{\circ}$ respectively. The black shaded region, green dashed line and blue solid line are similar to Fig .\ref{fig: J1843} calculated assuming $R_c=4$ kpc.}
\label{fig: J1959}
\end{figure} 
\begin{figure}[!h]
\centering
\includegraphics[scale=0.4]{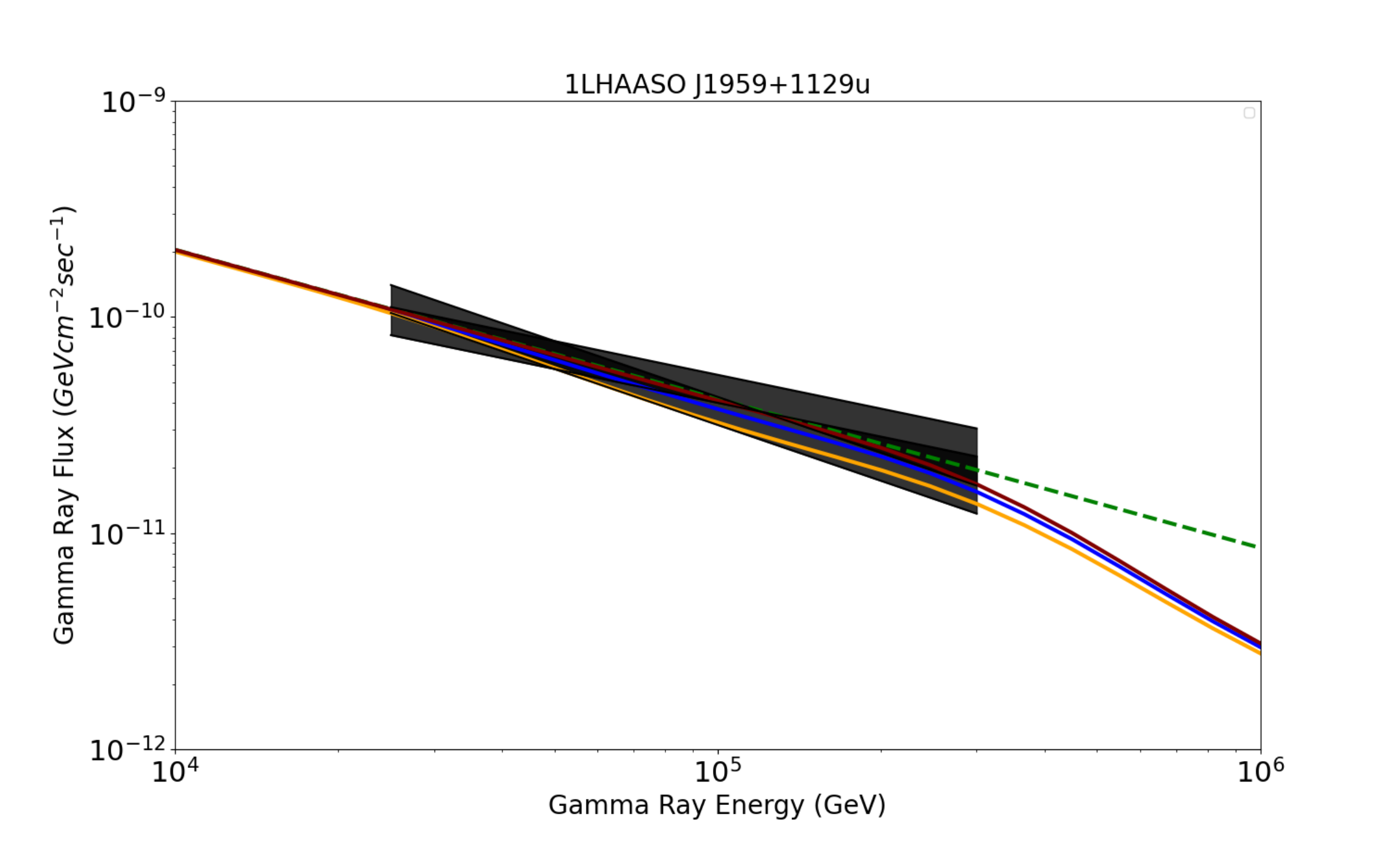}
\caption{ We have assumed three different radial dependencies of ISRF (maroon: $e^{-r/2}$; blue: $e^{-r/4}$ and orange: $e^{-r/8}$) to calculate the attenuated spectrum for the source in Fig \ref{fig: J1959}}
\label{fig: vary_rad_comp2}
\end{figure} 
In Fig \ref{fig: J1929} we have shown the gamma-ray spectrum of 1LHAASO J1929+1846u*. The attenuated spectrum has been calculated by varying the radial dependency of the ISRF, similar to the previous case. Fig \ref{fig: J1959} shows the gamma-ray spectrum of 1LHAASO J1959+1129u. Finally, in Fig \ref{fig: vary_rad_comp2}, we show the effect of the variation of the radial dependence of the ISRF on the attenuated spectrum of 1LHAASO J1959+1129u. Due to the large uncertainty in the measurement,  the attenuated spectra for all three cases ($e^{-r/2}$, $e^{-r/4}$ and $e^{-r/8}$) pass through the black shaded region. 

For all of these sources, more precise measurements of the UHE gamma-ray flux, which will reduce the uncertainty in flux measurement or the width of the black-shaded regions, would make it possible to constrain the radial component of ISRF more accurately along their directions from the observer on Earth. 

\par
We mentioned earlier that the intensity of ISRF also depends on longitude and latitude. Among many UHE gamma-ray sources, it is possible to select the ones with the least uncertainty in the measurement of UHE gamma-ray flux, which can constrain the radial dependence of ISRF in a particular direction.
 In Fig. 2. to Fig. 5. of \cite{2012A&A...545A..39R}, it is shown that the average intensity of ISRF is nearly constant for the values of longitude in the range of  5$^{\circ}$ to 30$^{\circ}$ and it drops slightly in the range of  40$^{\circ}$ to 50$^{\circ}$.
 In Table .1, we have listed some LHAASO sources in two groups whose longitudes are close to each other and whose latitudes are less than 1$^{\circ}$ so that the angular variation in the intensity of ISRF does not affect our result significantly.  We have calculated the values of $R_c$ allowed by the uncertainty in the measurement of UHE gamma-ray flux for these two groups of sources.  Between, 25$^{\circ}$ to 32$^{\circ}$  we get the maximum value of $R_c$ as 2 kpc and between 46$^{\circ}$ to 54$^{\circ}$  this values is 4 kpc for 1LHAASOJ1914+1150u and 1LHAASOJ1929+1846u. The maximum value of $R_c$ obtained using the UHE observational data from 1LHAASOJ1928+1813u is 6 kpc, as the uncertainty in the measurement of UHE gamma-ray flux is higher in this case.  
 This analysis can be extended to many more UHE gamma-ray sources if we can identify their possible counterparts because the distance is needed to calculate the optical depth.

\section{Discussion and Conclusion}
The Galactic structure is spatially complex with spiral arms, a central region dominated by a bulge/bar complex and warped stellar/dust disks.  There are different models of stars and dust distributions (\citet{2012A&A...545A..39R}, \citet{1998ApJ...492..495F}), which employ different spatial densities for both stars and dust but produce intensities very similar to observational data at near to far infrared frequencies in the solar neighbourhood \citep{Porter_2017}. Verifying the model predictions observationally in different parts of the Galaxy is important.
 \par
 In a recent paper by  \citet{bianchi2024revisitinglocalinterstellarradiation}, the local interstellar radiation field (LISRF) has been re-evaluated using the  {\it Gaia} Data Release 3 catalogue (DR3; \citet{2023A&A...674A...1G}) and new optical estimates from the {\it Pioneer} probes and also the nearby bright objects in the {\it Hipparcus} catalogue not included in DR3. The new LISRF is redder than the earlier estimate by  \citet{1983A&A...128..212M} and emits 30$\%$ more energy. The latest studies show that more investigations are needed to determine the ISRF distribution in the Galaxy.
 \par
 Gamma rays, having energy near a few hundred  TeV, mostly interact with dust emission, which is one of the components of ISRF and CMB radiation.
  We have shown how the scale length in the intensity profile of dust emission can be constrained with the UHE gamma-ray spectra. 
  We have shown the variation in the attenuated UHE gamma-ray spectra due to different values of the scale length in Fig.2. to Fig.7., assuming the intrinsic spectra follow a power law distribution. The scale length can be constrained by the uncertainty in the measurement of UHE gamma-ray flux. 
  We have constrained the value of the scale length $R_c$ in the radial distribution of dust emission using two groups of sources having latitudes less than 1$^{\circ}$. 
  The first group of sources has longitudes in the range of 25$^{\circ}$ to 32$^{\circ}$, and the second group of sources has longitudes in the range of 46$^{\circ}$ to 54$^{\circ}$. 
  Within these angular ranges, the effect of longitudinal variation in ISRF can be neglected. The scale length has a maximum value of 2 kpc for the first group and 4 kpc for the second group of sources. With more accurate measurements of the UHE gamma-ray spectra, it would be possible to get a more accurate estimate of $R_c$.
 With the detection of hundreds of UHE gamma-ray spectra extending to PeV energies with the ground-based gamma-ray observatories, it would be possible to study the variation in ISRF with radial distance, longitude and possibly latitude if there is a sufficient number of sources above and below the Galactic plane. This would help us to map the intensity distribution of the ISRF  in the Galaxy. The attenuation of the UHE gamma-ray spectrum near a few hundred TeV energies from the original power law spectrum is a unique feature to probe the ISRF as the attenuation is independent of the underlying mechanism of gamma-ray production or any other physical parameter of the sources. Future UHE gamma-ray telescopes can more precisely measure the UHE gamma-ray spectra, extending to higher energies.
 We look forward to more observational data with better statistics to extend this study in future. 
 
 \par
The detection prospects of very and UHE gamma-rays from extended sources with ASTRI Mini-Array (Astrofisica con Specchi a Tecnologia Replicante Italiana), CTAO (Cherenkov Telescope Array Observatory) and LHAASO have been discussed by  \citet{Celli_2024}. Due to their limited angular resolution, it is hard to identify the UHE gamma-ray sources with LHAASO and HAWC. Most sources have large angular extensions and are located among many other sources. Imaging Atmospheric Cherenkov Telescopes (IACTs) have better angular resolution, which will reduce source confusion. They also have a large Field of View (FoV), which will allow the next-generation telescopes like CTA and ASTRI Mini-Array
to resolve extended sources from crowded regions. Almost all the sources detected by  LHAASO-WCDA and in the H.E.S.S. Galactic plane survey will be within reach of ASTRI and CTA  with about 300 and 50 hours of exposure, respectively (\citet{Celli_2024}). Extensive Air Shower (EAS) and IACT measurements complement each other. IACTs are good for precise spectroscopic and morphological study of sources due to their excellent energy and angular resolution.  Hence, with ASTRI Mini-Array and CTA, it would be possible to measure the UHE gamma-ray spectra with high precision and determine the intensity of the ISRF along the line of sight, which attenuates the spectra.
\par
Although it is known that UHE gamma-ray spectra can constrain ISRF, to our knowledge, there is no other work as of now where the ISRF model parameter has been constrained with the LHAASO observed UHE gamma-ray spectra. Accurate measurements of UHE gamma-ray spectra by next-generation telescopes covering higher energy ranges will be particularly beneficial for further exploring the ISRF distribution and constraining the model parameters. 
\begin{deluxetable}{cccccc}
    \tablecaption{LHAASO Sources \label{tab:params}}
    \tablehead{
        \colhead{source} &
        \colhead{distance (kpc)} &
        \colhead{counterpart}&
        \colhead{longitude (degrees)}  & 
        \colhead{latitude (degrees)} &
        \colhead{ Range of Scale length ($R_c$ kpc)}   }
    \startdata
    \hline
    \hline
        1LHAASOJ1959+1129u& 9.6  & SNR G28.6-0.1  & 28.84   &  0.09 & {\bf $0 \leq R_c \leq 2.$} \\
        1LHAASOJ1837-0654u&6.6& SNR G24.7+0.6 & 25.21 &-0.08  &   {\bf $0 \leq R_c \leq  2.$}     \\
        1LHAASOJ1848-001u& 7.0& IGR J18490-0000& 32.61& 0.59&    {\bf $0 \leq R_c \leq  2.$}      \\      
 \hline
 1LHAASOJ1914+1150u& 14.01& PSR J1915+1150& 46.13&0.26&    {\bf $0 \leq R_c \leq   4.$ }        \\
 1LHAASOJ1929+1846u& 7.0& SNR G054.1+00.3& 53.88& 0.45&     {\bf $0 \leq R_c \leq 4.$}          \\
 1LHAASOJ1928+1813u&6.0&SNR G053.4+00.0& 53.28&0.42&  {\bf $0 \leq R_c \leq 6.$}                \\
 \hline
 \enddata
 \tablecomments{These sources have been taken from \citet{Cao_2024}. The radial dependence of ISRF has the scale factor $R_c$, whose values in various directions have been constrained by the uncertainty in the measurement of UHE gamma-ray flux.}
\end{deluxetable}

\section{Acknowledgement}
The author is thankful to Saikat Das and the referee for their helpful comments.
\bibliography{Reference}{}
\bibliographystyle{aasjournal}
\end{document}